# End-users needs and requirements for tools to support critical infrastructures protection

Michał Choraś and Rafał Kozik
Institute of Telecommunications
UTP Bydgoszcz
chorasm@utp.edu.pl

Rafał Renk and Witold Hołubowicz
Institute of Telecommunications. UTP Bydgoszcz
and
University of Adam Mickiewicz (UAM), Poznań

*Abstract*— The role of the services described in this paper is to support decisions in the Critical Infrastructure Protection (CIP) domain. Those services are perceived as the most fundamental functionalities, that will serve as a basis for the planned European simulation centre for modelling the behaviour of Critical Infrastructures (CI). The proposed services are: CI-related data accessing and gathering, threat forecasting and visualisation, consequence analysis, crowd management, as well as resources and capability management. In general, services proposed in the current paper will contribute to reducing the problem of overwhelming decision makers by too large amount of information. In the crisis, their decisions are made on the basis of the large amount of data related to the current situation, such as the status of CI, localisation of capabilities, weather and threat forecasts etc. The design of the services has been established with the help of the future end-users. The work presented in this paper is the result of preliminary activities performed in the FP7 project CIPRNet.

I. INTRODUCTION

According to [1] Critical Infrastructure (CI) can be described as asset, system or part thereof, which is essential for the maintenance of societal functions, health, safety, security, economic or social well-being of people, and the disruption or destruction of which would have a significant impact on a given country or region.

The importance of CI-related aspects stresses the fact that many research activities have been recently conducted in order to address different problems, including CI behaviour simulations [2], natural threat prediction and its impact evaluation [3], CI resilience [7], and cyber security of CI [8].

Protection of CI is a specific type of task. On the one hand, decisions taken for CIP purposes may impact human lives and material goods, threatened by both natural phenomena and as the consequence of human errors. On the other hand, such decisions must be taken in real-time – particularly during CI-related crisis. Most often, such decisions are taken by analysing a large amount of heterogeneous data.

In this paper, the services for CIP community and decision makers are presented to support decision making process in CIP, both in the preparedness ("cold") phase, as well as in the crisis ("hot") phase.

The goal of such services development is to increase the situational awareness of decision makers by extraction of the most necessary information from the large amount of heterogeneous data coming from different sources (such as real-time sensorial data).

Specification and development of the services proposed in this paper are the objectives of the Critical Infrastructure Preparedness and Resilience Research Network (CIPRNet) project – ongoing security research, co-funded by the European Commission's 7th Research Framework Program (FP7) [2].

In the first phase of the project, the end-users community was asked to express and share their needs and expectations related to increase of the effectiveness of modelling, simulation and CIP-related analysis environment and decision making process.

In this paper, the analysis of their requirements is provided, as well as the description of the demanded services which will later be designed in the CIPRNet project.

The rest of this paper is structured as follows:

- Section II includes the methodology and means used to collect end-user viewpoints in the CIPRNet project, the analysis of the end-user perspective drawn from this research and the summary of the key findings.
- In section III, the services to support CIP decision makers are presented. The proposed services include: data accessing and gathering, threat forecasting, threat visualisation, what-if and consequence analysis. Additionally, the services such as crowd management and resources and capability management are presented.
- In section IV, the general conclusions are presented.

II. END-USER NEEDS AND EXPECTATIONS

*A. CIP end-user views collection*

The identified stakeholders relevant to the project are representatives of the public, private, research and academia domains. The methods for gathering user views in the CIPRNet project were face-to-face meetings, remote user interviews and the CIPRNet questionnaire, filled in by the project end-users and domain experts. Outcomes of the collected questionnaires were a starting point in requirements specification process and in the specification of solutions described in this paper.

The questionnaire was designed in order to provide a broad view on current end-user problems, limitations and expectations, including big data issues. The most of the questions are open or semi-open. Therefore, respondents were neither limited in the expression of their opinions, nor biased by pre-defined options to choose.

Generally, the questionnaire has been divided into four blocks of questions, namely:

- General information about the respondents, particularly their organisations, range of activities, area in which he/she acts,
- Questions related to accessing the information, particularly concerning the availability of information about CI coming from private and public sectors and used during CI-related crisis,
- Questions about using decision support systems during respondent duties, providing information about decision support mechanisms and tools, their limitations, data exchange, standards, etc.,
- Questions about simulation and modelling for CI crisis management purposes.

The analysis of the CIPRNet questionnaires filled in by the CIPRNet end-users can be found in the next section.

*B. Questionnaire analysis*

Respondents who filled in the questionnaire are representatives of various organisations – from local and regional CI-related organisations to pan-European agencies, and from academic and applied researchers to CI operators. However, the majority of respondents are representatives of organisations that operate within nationwide range, and usually are from public emergency/crisis management centres.

The respondents assessed the availability of various information related to CI from various sectors and sources and gave them ratings. According to respondents ratings, generally there are no significant differences between the levels of availability of information, when comparing public and private sectors. The average ratings for public vs. private CI information availability (e.g. geo-localisation data, operational data and sensitive data about these infrastructures) are at the similar level. Considering the information about CI dependences, it is noticeable that such information during normal operation is significantly more easily accessible than during non-normal state of the CI functioning [3].

The questionnaire analysis shows that the hardest categories of information to be accessed include:

- Operational data of private sector CI,
- Information about CI across the national/regional borders,
- Information about CI across public-private sector borders,
- Information about CI dependencies during non-normal state.

In addition, respondents indicated that reliable data of CI financial aspects and CI failure status are also not easy to obtain from CI management entities.

According to end-users,

- climatic and weather information for specific (emergency) area, and
- geo-location information about public / private sector CI,

are described as the relatively easiest to obtain.

Concluding, most of the categories of information considered in the questionnaire (excluding e.g. the mentioned climatic/weather data) were assessed as relatively hard to obtain. This observation indicates a serious problem related to information accessibility, and what is worth noticing, challenges related to acquisition of necessary information exist regardless of the CI functioning sector (i.e. private versus public).

About 40% of respondents reported that they do not use any ICT-based support for their decisions. The majority of remaining 60% of respondents stated that they (or their organisations) use internally developed tools for specific purposes of their organisation, or alternatively, that they use various loosely coupled data sources (such as GIS resources, the weather data, etc.) to support decisions. Considering specific decision support tools (DSS) used by interviewees, examples such as C3M, IPCR or WebEOC have been listed. These systems are exploited for the crisis response planning, reporting, procedure and policy creating, resource allocation and tracking.

When asked about the analytical capabilities, as well as about usefulness and effectiveness of these systems during crisis-related decision-making, respondents presented different views. About half of them admitted that the used (decision support) systems do not meet their needs and that these systems are not tailored to the specific needs of their operation. As respondents emphasised, the main weakness of these systems is the need for advanced customisation (costly in terms of time, efforts, financing, etc.).

Other drawbacks include:

- lack of interconnectivity with the other systems (e.g. used by entities cooperating with stakeholder's organisation during CI-related crisis),

- lack of possibility to integrate the data from other entities/systems, hampering the cooperation between various organisations,
- limited capabilities of spatial visualisation of threats, and
- lack of capabilities to support comparison of the current situation to earlier forecasts.

Cross-border decision-making is another open gap of the used systems, impacting end-user operation.

The interviewees also listed various kinds of the information sources that are used for building the situational awareness in the emergency response efforts. These include mainly external sources such as cooperating entities and agencies involved in emergency response, which provide hydrological data, weather forecasts and the information about CI (including geo-location). Other sources of information are direct reports from the field/emergency area. Usually, such information is not publicly available. However, end-users can access that information in real-time or near real-time.

Respondents stated that the primary need for simulation models relates to consequences of CI object failure, employing e.g. cascade models of infrastructure failures. End-users indicated different scales of such consequences, varying from impact on another single system, up to consequences for national security, societal impact, national economy, etc.

Moreover, respondents noticed the lack of models supporting the estimation of CI restoration time, the identification of critical nodes (supporting CI objects prioritisation) and the simulation models relevant to a given, specific sector (e.g. applicable for health care services during CI failure).

Asked for the opinion on what should be improved in relation to decision-support for emergency management, respondents identified four main areas of interest:

1) Simulation and modelling, in particular the development of threat modelling and forecasting tools, e.g. for simulation of the consequences of possible decisions.

2) Estimation of crisis impact, both at a low level (e.g. impact of CI object failure on e.g. the hospital functioning), as well as at a higher level – for example estimation of CI failure costs, national economy losses, etc.

3) Emergency communication, namely:
- information/data sharing,
- timeliness of received information,
- exchange of information among cooperating agencies and organisations in real-time,
- compatibility of data formats,
- mechanisms to support informing about hazards, etc.

4) Cooperation and training between solution providers and emergency management teams.

According to the respondents, closer public-private cooperation also could improve the current situation in decision-making.

The respondents also indicated problems related to the current assessment of CI dependencies. The most significant examples include:

- Limited capabilities of simulations, particularly in terms of simulating interrelations between various CI and analysing the threats based on such relationships.
- Organisations and CI operators isolation. In other words, organisations often do not effectively take into account consequences of their infrastructure failures, exceeding beyond their organisations and impacting other sectors, companies, etc.
- Lack of systematic planning of CI protection and restoration after a crisis, as well as lack of procedures supporting such protection.
- Problems with identification of contact points that in the case of crisis should be immediately available for responsible entities.
- International standardisation in the CIP area.
- Information accessibility.
- Data validation and reliability.

*C. Key findings*

The end-users needs, expectations and requirements (presented in the previous subsection) can be categorised into the following aspects related to:

- decision support process,
- simulation and modelling,
- access to the real-time data and critical information.

The key findings that have been identified after the analysis of the mentioned aspects are as follows:

1) End-users expect more advanced, customised and tailored (to their needs) decision support solutions, which will allow for flexible spatial threats visualisation, easy integration with new data sources or other systems, and information sharing between different entities engaged into a crisis management process.

2) End-users lack accurate models and simulation tools that will allow for consequences, impact and risk analysis of CI failures and cascading effects. The forecasting capabilities are emphasised as one of the most desired.

3) End-users articulated the need of access to information related to CI from various sectors. They emphasised difficulties in gaining data related to operational state of private sectors CI and CI-related information across public-public and national-regional borders.

## III. CIPRNET SERVICES

In order to cover the key findings (described in the previous section) coming from the end-user perspective analysis, the following decision support services have been specified:

- Consequence analysis,
- Threat forecasting,
- Threat visualisation,
- Data accessing and gathering,
- Crowd management / Crowd mapping,
- Resources and capability management.

It is expected that the functionalities and the number of services provided by the DSS will evolve over time. Therefore, a plug-and-play and easy to extend architecture of the DSS is anticipated (see Fig. 1).

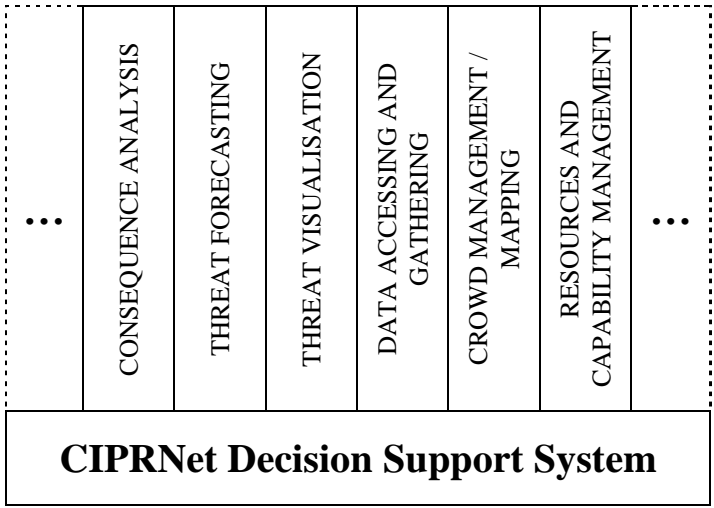

Fig. 1. Extendable DSS architecture

The proposed DSS will have two distinctive operational modes, namely "Hot Phase" and "Cold Phase" decision support (Fig. 2).

The "Cold Phase" is computationally intensive, therefore, it is dedicated to postmortem analysis and CI operators training purposes. Among others it will heavily rely on historical data, modelling, simulation and analysis (MS&A).

The "Hot Phase" includes continuous and real-time risk assessment, threat forecasting and consequence analysis conducted using real-time data during the real crisis. These aspects are explained in the next sections.

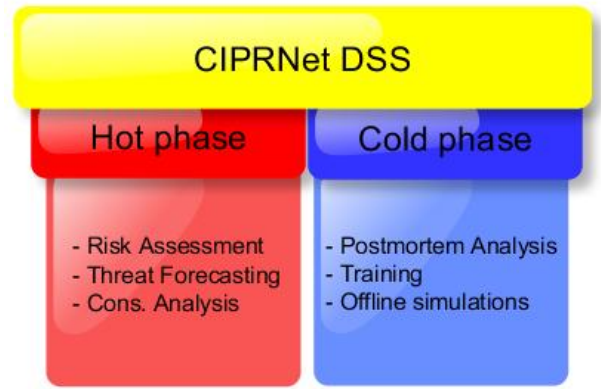

Fig. 2. Operational modes of the CIPRNet DSS

### A. Data accessing and gathering

The data accessing and gathering service allows for collecting and storing the data relevant to decisions taken for CIP purposes. The data is stored in dedicated database. The purpose of this service is to provide all input data, necessary to support the decisions by the CIPRNet DSS. Particularly, the service provides the data used by the threat forecasting service.

The data stored in the database is structured, based on pre-defined layered model, including territorial, socio-economical, infrastructure, and historical data. Additionally, each layer can be divided into further sub-layers.

The sources of data can be e.g. governmental repositories, infrastructure operators and results of modelling and simulation activities.

### B. Threat forecasting

The threat forecasting service is one of the key features of the DSS. This service provides the DSS with capability to forecast natural phenomena that can cause physical damages of CI and that can impact the normal operation of CI. Such phenomena include e.g. heavy rain, flooding, landslide, drought, heat wave, etc.

The threat forecasting service is designed in modular way. Each module being an element of the service is dedicated to forecasting a specific phenomenon. Particular modules are interconnected with the corresponding layers of "data accessing and gathering" service. Each module uses data coming from the specific layer to run the appropriate model and to forecast specific threats on a specific area.

### C. Threat visualisation

This service is intended to provide DSS user the various visualisation capabilities. The role of this service is to use different means to visualise a variety of aspects, that may influence decision making process for CIP purposes. Therefore, visualisation service is oriented on the usability of the DSS, minimising the amount of data delivered to the decision makers and changing the form of these data from e.g.

tabular, raw data into more understandable and user friendly graphic presentation. Examples of the visualisation capabilities may include presentation of predicted CI affecting threats in aerial or GIS maps, as well as visualisation of damage level, that can be caused by a natural phenomenon. Threat visualisation service is strongly interconnected with other DSS services proposed in this paper, since the visualised data are provided e.g. by "data accessing and gathering" service and processed by the threat forecasting service. The main requirements that this service must satisfy include efficient and flexible access to the threat visualisation data, the ability to handle multiple simultaneous requests for visualisation and the ability to share provided visualisations among private and public stakeholders, emergency managers and common citizens involved in disaster response.

*D. Consequence analysis*

The consequence analysis is the service included in the CIPRNet DSS, that offers the added value to the decision making process. This service enables decision-makers and operators to analyse the impact of natural hazards on CI failures and to examine their possible short and long term consequences. The consequence analysis will rely on both real-time (sensorial, e.g. geo-seismic) data, as well as statistical and historical data about the past, similar incidents. Real-time status information about the functioning of a particular CI element and meteorological data are also involved in the consequence analysis.

*E. What-if analysis*

The "What-if analysis" is one of the CIPRNet services with the main goal to provide the end-user with the simulation capabilities, which allow CI-related aspects to be investigated. Among others, the "What-if analysis" will provide the end-user with tools, which will allow them to analyse different crisis scenarios that may affect critical infrastructures. The analysis will allow the end-user to investigate different courses of actions and to evaluate their consequences. The core functionalities of this service will be enabled with tools and frameworks for federated simulation [4]. The underpinnings for this have been established by DIESIS [5][6] project, of which CIPRNet is the successor.

*F. Crowd management / Crowd mapping*

The crowd mapping service aims at large scale sharing of user (citizen)-related information based on their geo-positioning data. Input data for the crowd mapping service is data coming from a variety of sources like mobile device or web pages, logging the physical localisation of a user. The main added value for the decision making process regarding crowd management for CIP purposes is support for evacuation management during a crisis situation. The main functionalities of this service include real-time human tracking during a given crisis situation and in a specific area, evacuation route analysis and communication with citizens established based on citizen geo-position.

*G. Resources and capability management*

The capability management service is designed in a way that allows the DSS operator to automatically match the capabilities and resources that crisis management entities have at their disposal to the relevant crisis circumstances. It will increase the chance for prevention of negative impact of crisis situation and for timely intervention during crisis.

The main service functionalities are:

- Identification of means (resources) and sensors that are best suited to react in the current crisis situation,
- Prioritisation of the identified means, taking into account specific limitations, type of CI failure and the properties of the identified capabilities (e.g. availability, reliability, time-to-deploy, effectiveness in a given situation, restoration time, etc.),
- Visualisation of the geographical dependencies between the available capabilities and the area affected by the crisis.

IV. CONCLUSIONS

In this paper we described the set of services composing the decision support system for Critical Infrastructure Protection (CIP).

These services are perceived as the most fundamental and were articulated by the end-users. Those functionalities will serve as the basis for the planned European simulation centre for modelling the behaviour of Critical Infrastructures.

The proposed services add value to current CIP efforts and are designed in a way that will improve effectiveness of CIP decision making. The proposed services are: data accessing and gathering, threat forecasting and visualisation, consequence analysis, crowd management, as well as resources and capability management.

Particularly, the threat visualisation, crowd mapping and resources (capability) management services will allow DSS users to deal with constantly changing information during CI crisis and to more effectively build a reliable view on a crisis situation. This will be achieved through a more clear and better tailored to user needs form of information presentation, including geographical representation of crowd sources, visualisation of CI elements, CI affecting threats and means for mitigating them in GIS-based and satellite maps.

In addition, the consequence analysis service will allow users to examine various crisis scenarios and to learn about the results of various possible decisions. The data accessing and gathering service is designed to provide only the most relevant and necessary data to particular DSS services, launched for specific purposes. Therefore, this service will also reduce the problem of overwhelming decision makers by too large amount of information.

Concluding, the services described in this paper are proposed to support decisions for CIP purposes. Such decisions are always influenced by the large amount of information about the current situation, therefore the proposed services are the CIPRNet response to the big data problem, that CI decision makers must face.

ACKNOWLEDGMENT

This work is partly funded by the European Commission under grant number FP7-312450-CIPRNet. The support is gratefully acknowledged.